\documentclass[11pt,letter,floatfix]{article}
\pdfoutput=1
\usepackage{jcappub}
\usepackage{float}
\usepackage{amssymb}

\usepackage{gensymb}
\usepackage{caption}
\usepackage{graphicx}
\usepackage{amsmath}
\usepackage{wasysym}
\usepackage{verbatim}
\usepackage[utf8]{inputenc}
\usepackage{adjustbox}
\usepackage{graphicx}

\usepackage{lineno}
\begin{document}
\title{An optimized search for dark matter in the galactic halo with HAWC}
\author[a]{A.~Albert}

\author[b]{R.~Alfaro}

\author[c]{C.~Alvarez}

\author[d]{J.C.~Arteaga-Velázquez}

\author[b]{D.~Avila Rojas}

\author[e]{H.A.~Ayala Solares}

\author[b]{E.~Belmont-Moreno}

\author[c]{K.S.~Caballero-Mora}

\author[f]{T.~Capistrán}

\author[g]{A.~Carramiñana}

\author[h]{S.~Casanova}

\author[i]{O.~Chaparro-Amaro}

\author[d]{U.~Cotti}

\author[j]{J.~Cotzomi}

\author[k]{E.~De la Fuente}

\author[g]{R.~Diaz Hernandez}

\author[a]{B.L.~Dingus}

\author[l]{M.A.~DuVernois}

\author[a]{M.~Durocher}

\author[k]{J.C.~Díaz-Vélez}

\author[b]{C.~Espinoza}

\author[m]{K.L.~Fan}

\author[f]{N.~Fraija}

\author[n]{J.A.~García-González}

\author[f]{F.~Garfias}

\author[f]{M.M.~González}

\author[m]{J.A.~Goodman}

\author[a]{J.P.~Harding}

\author[o]{D.~Huang}

\author[c]{F.~Hueyotl-Zahuantitla}

\author[f]{A.~Iriarte}

\author[p]{V.~Joshi}

\author[a]{G.J.~Kunde}

\author[q]{J.~Lee}

\author[b]{H.~León Vargas}

\author[r]{J.T.~Linnemann}

\author[f]{A.L.~Longinotti}

\author[s]{G.~Luis-Raya}

\author[r]{J.~Lundeen}

\author[a]{K.~Malone}

\author[j]{O.~Martinez}

\author[i]{J.~Martínez-Castro}

\author[t]{J.A.~Matthews}

\author[j]{E.~Moreno}

\author[e]{M.~Mostafá}

\author[h]{A.~Nayerhoda}

\author[u]{L.~Nellen}

\author[r]{A.~Peisker}

\author[s]{E.G.~Pérez-Pérez}

\author[q]{C.D.~Rho}

\author[g]{D.~Rosa-González}

\author[j]{H.~Salazar}

\author[r]{D.~Salazar-Gallegos}

\author[b]{A.~Sandoval}

\author[b]{J.~Serna-Franco}

\author[v]{R.W.~Springer}

\author[s]{O.~Tibolla}

\author[r]{K.~Tollefson}

\author[g]{I.~Torres}

\author[w]{R.~Torres-Escobedo}

\author[o]{R.~Turner}

\author[g]{F.~Ureña-Mena}

\author[j]{L.~Villaseñor}

\author[o]{X.~Wang}

\author[w]{H.~Zhou}

\author[d]{C.~de León}

\affiliation[a]{Physics Division, Los Alamos National Laboratory, Los Alamos, NM, USA }

\affiliation[b]{Instituto de F\'{i}sica, Universidad Nacional Autónoma de México, Ciudad de Mexico, Mexico }

\affiliation[c]{Universidad Autónoma de Chiapas, Tuxtla Gutiérrez, Chiapas, México}

\affiliation[d]{Universidad Michoacana de San Nicolás de Hidalgo, Morelia, Mexico }

\affiliation[e]{Department of Physics, Pennsylvania State University, University Park, PA, USA }

\affiliation[f]{Instituto de Astronom\'{i}a, Universidad Nacional Autónoma de México, Ciudad de Mexico, Mexico }

\affiliation[g]{Instituto Nacional de Astrof\'{i}sica, Óptica y Electrónica, Puebla, Mexico }

\affiliation[h]{Institute of Nuclear Physics Polish Academy of Sciences, PL-31342 IFJ-PAN, Krakow, Poland }

\affiliation[i]{Centro de Investigaci\'on en Computaci\'on, Instituto Polit\'ecnico Nacional, M\'exico City, M\'exico.}

\affiliation[j]{Facultad de Ciencias F\'{i}sico Matemáticas, Benemérita Universidad Autónoma de Puebla, Puebla, Mexico }

\affiliation[k]{Departamento de F\'{i}sica, Centro Universitario de Ciencias Exactase Ingenierias, Universidad de Guadalajara, Guadalajara, Mexico }

\affiliation[l]{Department of Physics, University of Wisconsin-Madison, Madison, WI, USA }

\affiliation[m]{Department of Physics, University of Maryland, College Park, MD, USA }

\affiliation[n]{Tecnologico de Monterrey, Escuela de Ingenier\'{i}a y Ciencias, Ave. Eugenio Garza Sada 2501, Monterrey, N.L., Mexico, 64849}

\affiliation[o]{Department of Physics, Michigan Technological University, Houghton, MI, USA }

\affiliation[p]{Erlangen Centre for Astroparticle Physics, Friedrich-Alexander-Universit\"at Erlangen-N\"urnberg, Erlangen, Germany}

\affiliation[q]{University of Seoul, Seoul, Rep. of Korea}

\affiliation[r]{Department of Physics and Astronomy, Michigan State University, East Lansing, MI, USA }

\affiliation[s]{Universidad Politecnica de Pachuca, Pachuca, Hgo, Mexico }

\affiliation[t]{Dept of Physics and Astronomy, University of New Mexico, Albuquerque, NM, USA }

\affiliation[u]{Instituto de Ciencias Nucleares, Universidad Nacional Autónoma de Mexico, Ciudad de Mexico, Mexico }

\affiliation[v]{Department of Physics and Astronomy, University of Utah, Salt Lake City, UT, USA }

\emailAdd{lundeenj@msu.edu}
\emailAdd{jpharding@lanl.gov}
\emailAdd{tollefs2@msu.edu}
\date{\today}
\abstract{
The Galactic Halo is a key target for indirect dark matter detection. The High Altitude Water Cherenkov (HAWC) observatory is a high-energy ($\sim300$ GeV to $>100$ TeV) gamma-ray detector located in central Mexico. HAWC operates via the water Cherenkov technique and has both a wide field of view of ~2 sr and a \textgreater 95\% duty cycle, making it ideal for analyses of highly extended sources. We made use of these properties of HAWC and a new background-estimation technique optimized for extended sources to probe a large region of the Galactic Halo for dark matter signals. With this approach, we set improved constraints on dark matter annihilation and decay between masses of 10 and 100 TeV. Due to the large spatial extent of the HAWC field of view, these constraints are robust against uncertainties in the Galactic dark matter spatial profile. 
}
\maketitle

\section{Introduction}
\label{sec:intro}
The mass of the known universe is dominated by a dark component that is not optically observable.  First observed by Fritz Zwicky's 1930s study of the Coma Cluster, evidence of this dark matter has gradually accumulated over the years.  Its effects can be seen in phenomena such as galactic velocity dispersion and gravitational lensing through galaxy clusters \cite {andernach2017english,dark_evidence}.   These effects cannot be explained by normal, luminous matter given the known laws of gravity, leading to the hypothesis that halos of dark matter surround the luminous components, which provides the mass necessary to explain these observations \cite{wimp}.  Current estimates place the dark matter contribution to the mass of the universe at 86\% \cite{planck}.   However, the composition of the dark matter remains unknown.

One of the most popular candidates are Weakly Interacting Massive Particles, or WIMPs.  As indicated by the name, these are hypothesized particles with a non-zero mass that interact via a weak-scale force and, in many theoretical frameworks, are their own anti-particle \cite{wimp,wimpproceeding}.  Assuming WIMPs were in thermal equilibrium in the early Universe, this weak-scale self-annihilation naturally reproduces the observed relic density while maintaining the cold dark matter state of the known universe \cite{planck, dark_back}.  WIMPs are therefore one of the most promising dark matter candidates and many contemporary experiments aim to find evidence of WIMP interactions.

WIMP annihilation and decay can produce Standard Model particles through weak interactions. These particles will, in turn, produce gamma-ray photons mainly via pion decay, but also through inverse Compton scattering of photons off produced electron-positron pairs \cite{patproceeding}. The expected differential gamma-ray flux, $d\Phi$, per unit energy, $dE$, from a dark matter halo is described by the following equations modeling annihilation and decay, respectively:
\begin{equation}
\frac{d\Phi}{dE}_{\mathrm{ann}}=\frac{J \langle \sigma v \rangle}{8 \pi M^2}\frac{dN(M,channel)}{dE} \enspace ,
\label{dm-ann}
\end{equation}
\begin{equation}
\frac{d\Phi}{dE}_{\mathrm{decay}}=\frac{D}{4 \pi M \tau}\frac{dN(M,channel)}{dE} \enspace ,
\label{dm-dec}
\end{equation}
where $\langle \sigma v \rangle$ is the velocity-weighted annihilation cross section, $\tau$ is the decay lifetime, $M$ is the dark matter particle mass, and the $J$ and $D$ factors contain the astrophysical information of the assumed dark matter density profile.  See ref.~\cite{patproceeding} for a detailed derivation of these functional forms.  

The $J$-factor and $D$-factor are defined, respectively, as:
\begin{equation}
J= \int\int \rho_{dm}^2(l,\Omega) dl d\Omega \enspace,
\label{j-fac}
\end{equation}
\begin{equation}
D= \int\int \rho_{dm}(l,\Omega) dl d\Omega \enspace ,
\label{d-fac}
\end{equation}
where $\rho_{dm}$ is the dark matter density profile (typically containing both a smooth component and a contribution from sub-halos), $dl$ is the line of sight to the halo, and  $\Omega$ is the solid angle of the observation.  The J-factor contains a density-squared factor due to annihilation requiring two particles to interact, while the D-factor includes only one power of the density profile since decay is a single-particle process.

The quantity $\frac{dN}{dE}$ in eq.~\ref{dm-ann} and eq.~\ref{dm-dec} is the gamma-ray spectrum from a single dark matter interaction \cite{dark_back}. This analysis uses spectra computed with the Poor Particle Physicist's Cookbook (PPPC) model as derived by Cirelli et al. \cite{pppc}, which include corrections arising from the electroweak coupling. 


\section{The Galactic Halo}
\label{sec:gh}

The Galactic Halo of the Milky Way galaxy is expected to yield an extremely high dark matter gamma-ray flux and is a promising region for probing WIMP signals. 
The exact behavior of the Galactic Halo density profile is not well constrained towards the center, so we will consider different parameterizations consistent with numerical simulations and observational data.  

Many fits to numerically simulated halos consisting only of dark matter favor the Einasto density profile, which is characterized by a sharp cusp towards the halo center \cite{einasto3, einasto1, einasto2} and given by:
\begin{equation}
\rho(r)=\rho_{s} e^{\frac{-2}{\alpha}[(r/r_{s})^{\alpha}-1]} \enspace ,
\label{einastogc}
\end{equation}
where $\rho_{s}$ is a normalization constant on the dark matter mass density determined by the total halo mass, $r_{s}$ is the characteristic scale radius of the halo, and $\alpha$ determines the profile's curvature, which is fixed to a value of 0.17 for the Galactic Halo \cite{clumpy}.
Observations of dark matter halos favor a Burkert density profile \cite{burkert}, which lacks the center cusp and instead flattens towards the center (referred to as being ``cored" in contrast to ``cuspy"). The Burkert profile is given by:  
\begin{equation}
\rho(r)=\frac{\rho_{s}}{(1+r/r_{s})(1+(r/r_{s})^{2})} \enspace ,
\label{burkertgc}
\end{equation}
where $\rho_{s}$ and $r_{s}$ are again the density normalization and scale radius, respectively. Additionally, more recent N-body simulations that include baryonic matter favor the cored shape parameterized by eq.~\ref{burkertgc} \cite{cusp_flatteing}.  

\begin{figure}
	\centering
	\includegraphics[width=0.7\textwidth]{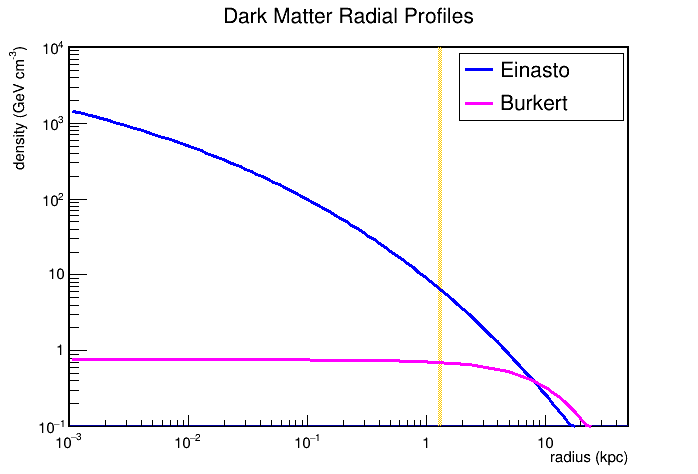}
	\caption{A comparison of the dark matter density profile behavior as a function of distance from the Galactic Halo center.  The Einasto (cuspy) profile \cite{einasto3, einasto1, einasto2} differs by many orders of magnitude from the Burkert (cored) profile \cite{burkert} towards the center, but by less than an order of magnitude for the region considered in this analysis (to the right of the yellow line).}
	\label{fig:profiles-comparison}
\end{figure}

The behavior of the different density profiles is illustrated in figure~\ref{fig:profiles-comparison} for the Galactic Halo.  Towards the center, the profile behavior diverges considerably.  Therefore, any expected dark matter gamma-ray flux computed using only the region close to the center will have large systematic uncertainties (over six orders of magnitude) arising from the choice of density profile. However, an experiment with a large instantaneous field of view can observe a larger portion of the Galactic Halo and is therefore more robust, as will be explained in section ~\ref{sec:hawc}. For the particular declination range of this analysis, only galactic radii \textgreater1.27 kpc are considered, where the Burkert and Einasto profiles are less than an order of magnitude apart. The simulated values of the dark matter profile considered in this paper are given in Table~\ref{table:profileparams}.

\begin{table}
	\begin{center}
		\begin{tabular}{|c|c|c|c|}
			\hline
			$R_{\odot}$ (kpc) & $\rho_{\odot}$ ($\mathrm{GeV/cm^{3}}$)  &  $r_s$ (kpc)   &  $\alpha$    \\
			\hline
			8 &  0.4 & 15.7 & 0.17  \\  
			\hline
		\end{tabular}
	\end{center}
	\caption{Parameters used in the assumed dark matter density profiles (eq. \ref{einastogc} and eq. \ref{burkertgc}). $R_{\odot}$ and $\rho_{\odot}$ are the distance from the Sun to the Galactic center and the local dark matter density of the solar system, respectively, which are used to define the normalization for the density profiles ($\rho_s$ in eq.~\ref{einastogc} and eq.~\ref{burkertgc}). The scale radius, $r_s$, is chosen based on the value from the Aquarius simulation, as is the value of $\alpha$ for the Einasto profile~\cite{Navarro:2008kc}. }
	\label{table:profileparams}
\end{table}



\section{The HAWC detector}
\label{sec:hawc}

The High Altitude Water Cherenkov (HAWC) detector is a gamma-ray observatory located on the side of the Sierra Negra volcano in Mexico. HAWC observes particle-induced air showers using an array of 300 Water Cherenkov Detectors (WCDs) and covers an area of 22,000 $\mathrm{m^{2}}$. It is one of the most sensitive instruments for measuring multi-TeV gamma rays currently operating. 
HAWC is sensitive to gamma rays with energies of at least 300 GeV and up to over 100 TeV, and is therefore well-suited for detecting possible signals from multi-TeV-mass WIMPs.  In addition, HAWC operates on a near-continuous duty cycle with a roughly 2 sr instantaneous field of view that makes it ideal for performing survey-style observations \cite{HAWC:2020hrt}. See ref.~\cite{hawcback} for details on the standard HAWC reconstruction and binning techniques used in this analysis and ref.~\cite{historical:2023opo} for more details on the HAWC design.

HAWC uses a forward-folding approach to fit the true energy spectrum of a source from the binned data, taking into account the HAWC detector response obtained from simulation.  Events in this analysis are primarily binned by the fraction of available PMTs triggered, which serves as an energy estimator as detailed in ref.~\cite{hawcback}.  Events are further categorized into spatial bins (also referred to as ``pixels" in later sections) using the \textsc{heal}pix pixelization scheme \cite{healpix}.  
The best-fit spectrum for a given source is calculated by using a maximum-likelihood approach as detailed in refs.~\cite{hawcback} and \cite{dwarf}.  The statistical significance is then computed using the test statistic (TS) defined as:
\begin{equation}
TS = 2 \log\big(L_{\mathrm{max}}/L_0\big),
\label{ts}
\end{equation} 
where $L_{\mathrm{max}}$ is the maximum likelihood resulting from the fit, and $L_0$ is the likelihood for a background-only model (where the expected signal is zero in all bins).

Since the background-only hypothesis is contained entirely within the signal-plus-background hypothesis, Wilks' theorem can be used to interpret the TS.  Wilks' theorem shows that if the background-only hypothesis is true, in the high-statistics limit, the TS will follow a $\chi^2$ distribution with degrees of freedom equal to the difference in the degrees of freedom between the signal hypothesis and background-only hypothesis \cite{wilks}.  The binning scheme used in this analysis has been previously shown to contain sufficient statistics to apply Wilks' Theorem in ref.~\cite{hawcback} and \cite{HAWC:2020hrt}. Therefore, the appropriate $\chi^2$ distribution for a given signal hypothesis can be used to transform the TS into a measure of the significance by which the background-only model is rejected.

Due to its wide field of view, HAWC can probe a large region of the Galactic Halo for dark matter gamma-ray signals.  Using a large region of interest (ROI) farther away from the Galactic Center substantially mitigates the density profile systematic, since the cored and cuspy profiles do not differ as strongly.  In the cored case, the sensitivity lost from the lack of an assumed central cusp is also mitigated by the larger ROI, allowing more flux to be observed over a larger area. 


\section{Large-scale background estimation} 
\label{sec:bkg}

The majority of air showers detected by HAWC come from charged cosmic rays. Gamma-ray and cosmic-ray events can be separated by the different characteristics of the lateral charge distributions they create across the array.  Using gamma-hadron discrimination cuts to quantify this effect, HAWC is able to reject greater than 99\% of the hadronic background above $\sim$3 TeV \cite{hawcback}.  However, due to the extremely high relative abundance of charged cosmic rays to gamma rays, a substantial amount of background remains even after cuts.

This analysis estimates the background using a technique that is optimized for highly extended source hypotheses \cite{pooja_thesis}.  The technique uses two datasets; one with standard gamma-hadron cuts applied, and another with reversed cuts meant to pass hadrons, thereby creating a dataset consisting primarily of cosmic rays.  This approach then attempts to map the behavior of the pure cosmic-ray maps back to the distribution of hadronic background in the standard maps.  A series of calculations are then performed to derive what is called the $\alpha$-factor (analogous to the exposure factor used by Li and Ma \cite{li-ma})\footnote{In order to be compatible with usage in the relevant literature, we are using the symbol $\alpha$ both for this quantity, and the Einasto parameter in eq 2.1; the meaning should be from the context.}.  This $\alpha$-factor is used to compute the estimated hadronic background in the $i$th bin of the standard maps, where $i$ runs over both spatial and analysis (energy) bins. This technique is only applied to events where more than 16.2\% of the HAWC array triggers due to known  biases for smaller events. The $\alpha$-factor is used to compute the estimated background in a given pixel, $N^{\mathrm{BKG}}_i$, via the following equation:
\begin{equation}
N^{\mathrm{BKG}}_i = \alpha_i \times H_i \enspace ,
\end{equation}
where $H_i$ is the content of the hadron maps in a given pixel and $\alpha_i$ is the pixel-by-pixel computed $\alpha$-factor.

To compute $\alpha_i$, it is decomposed into the RA-dependent and Dec-dependent terms (see ref.~\cite{pooja_thesis}) as follows:
\begin{equation}
\alpha_i(\mathrm{RA,Dec}) = a_i(\mathrm{Dec}) \times b_i(\mathrm{RA}) \enspace ,
\label{alpha}
\end{equation}
where the $i$ index runs over spatial pixels.  Assuming that each component is independent of the other,  the $a_i(\mathrm{Dec}) $ component is calculated by:
\begin{equation}
a_i(\mathrm{Dec})  = \frac{G_{\mathrm{Dec}}}{H_{\mathrm{Dec}}} \enspace ,
\end{equation}
where $G_{\mathrm{Dec}}$ and $H_{\mathrm{Dec}}$ are the counts averaged within a 0.5 degree window around the declination of a given pixel in the gamma and hadron maps, respectively.  Since the HAWC event rate is highly dependent on declination, the 0.5 degree averaging window was chosen to be on the order of the size of a single spatial pixel to minimize mixing of counts between different declinations.  The RA-dependence, $b_i(\mathrm{RA})$, is expected to depend on the pixel-by-pixel count ratio, $\frac{G_i}{H_i}$, but with the declination dependence factored out.  The equation is then:
\begin{equation}
b_i(\mathrm{RA}) = S_G\bigg(\frac{G_i}{H_i} \frac{1}{a_i(\mathrm{Dec})}\bigg) \enspace ,
\end{equation}
where $S_G$ is a Gaussian smoothing function with width $\sigma$ (in degrees of RA). 
This is repeated for each analysis (energy) bin, where the value of the smoothing function is determined by the analysis bin index $k$ as $\sigma = 2k+15$, where k ranges from 3 to 9.  Since the average count in each map falls off in the higher energy bins, the smoothing width is made wider in these bins to gather additional statistics.  See ref.~\cite{pooja_thesis} for further details.

Note that the $\alpha$-background technique reveals a previously-undetected excess just to
the north side of the Galactic Plane, as shown in ref.~\cite{pooja_thesis}. This new excess is morphologically inconsistent with gamma rays from dark matter emission originating from the main Galactic Halo and we therefore include it as a part of the list of astrophysical sources and excesses to be removed when the dark matter search is performed.
\footnote{The collaboration is currently looking into the nature of this gamma-ray source. However, to be conservative we have included this in the masked exclusion region of the analysis presented here.}


\section{Analysis method}
\label{sec:method}

\subsection{Region of interest selection}
\label{subsec:roi}

In order to estimate the ROI for this analysis, we combine the characteristic sensitivity of HAWC from ref.~\cite{all-sky} with the simulations of the Galactic Halo dark matter density profile. We use the \textsc{clumpy} software package \cite{clumpy} to generate these simulations with input parameters given in table.~\ref{table:profileparams}.

\begin{figure}
	\includegraphics[width=\textwidth]{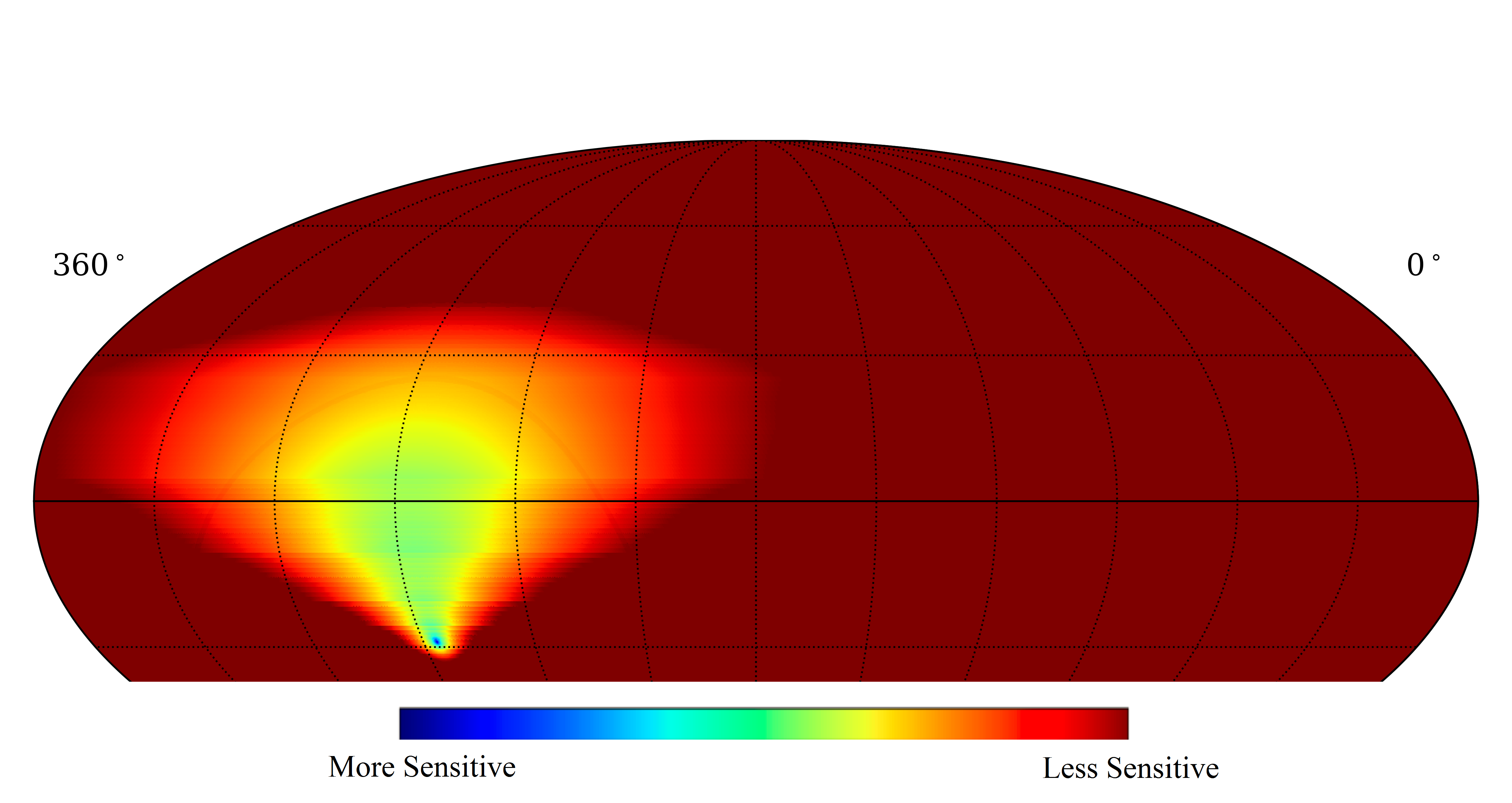}
	\includegraphics[width=\textwidth]{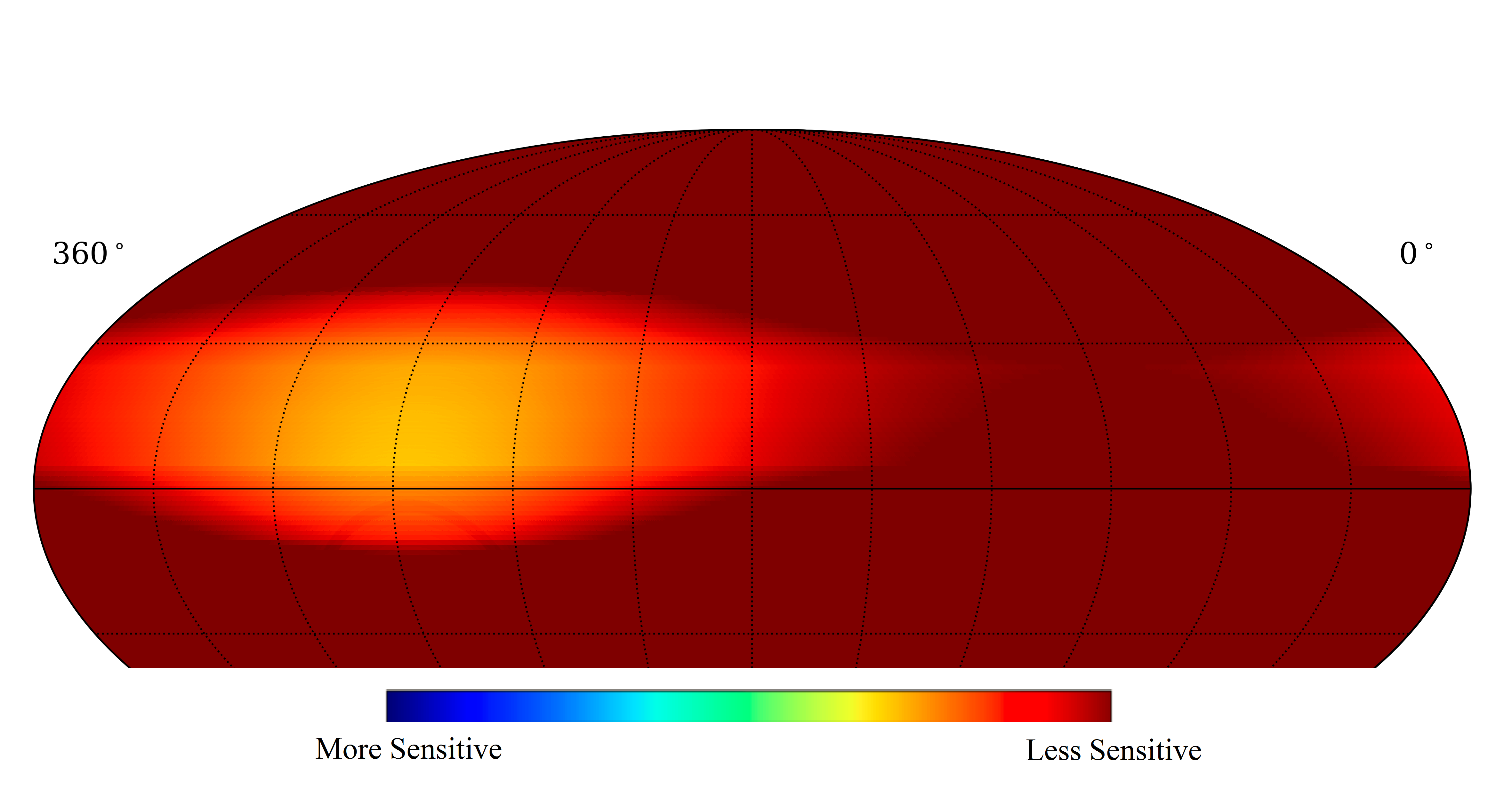}
	\caption{Relative sensitivity of each spatial bin to a dark matter gamma-ray signal, plotted as a function of RA and Dec. Brighter coloring indicates more sensitive regions. The top figure assumes an Einasto spatial profile while the bottom figure assumes a Burkert spatial profile. Points more sensitive to dark matter will be included in the regions of interest for the full analysis and excluded from the background estimation.}
	\label{fig:sensitivity_map}
\end{figure}

The $\alpha$-background approach requires removing spatial bins expected to contain excess from the background-estimation calculations.  The optimal ROI may, in general, depend on the choice of density profile.  Since cuspy profiles peak sharply towards the Galactic center, additional sensitivity could be gained by considering pixels at the extreme southern edge of the HAWC field of view.  In contrast, cored profiles are expected to derive almost all of their sensitivity from regions overhead, since the increase in expected flux towards the center is not enough to dominate loss of sensitivity at high zenith angles. The resulting plots are shown from both a cuspy (Einasto) and cored (Burkert) profile in figure~\ref{fig:sensitivity_map}, where lighter colors indicate pixels more favorable for DM searches.

\begin{figure}
	\includegraphics[width=\textwidth]{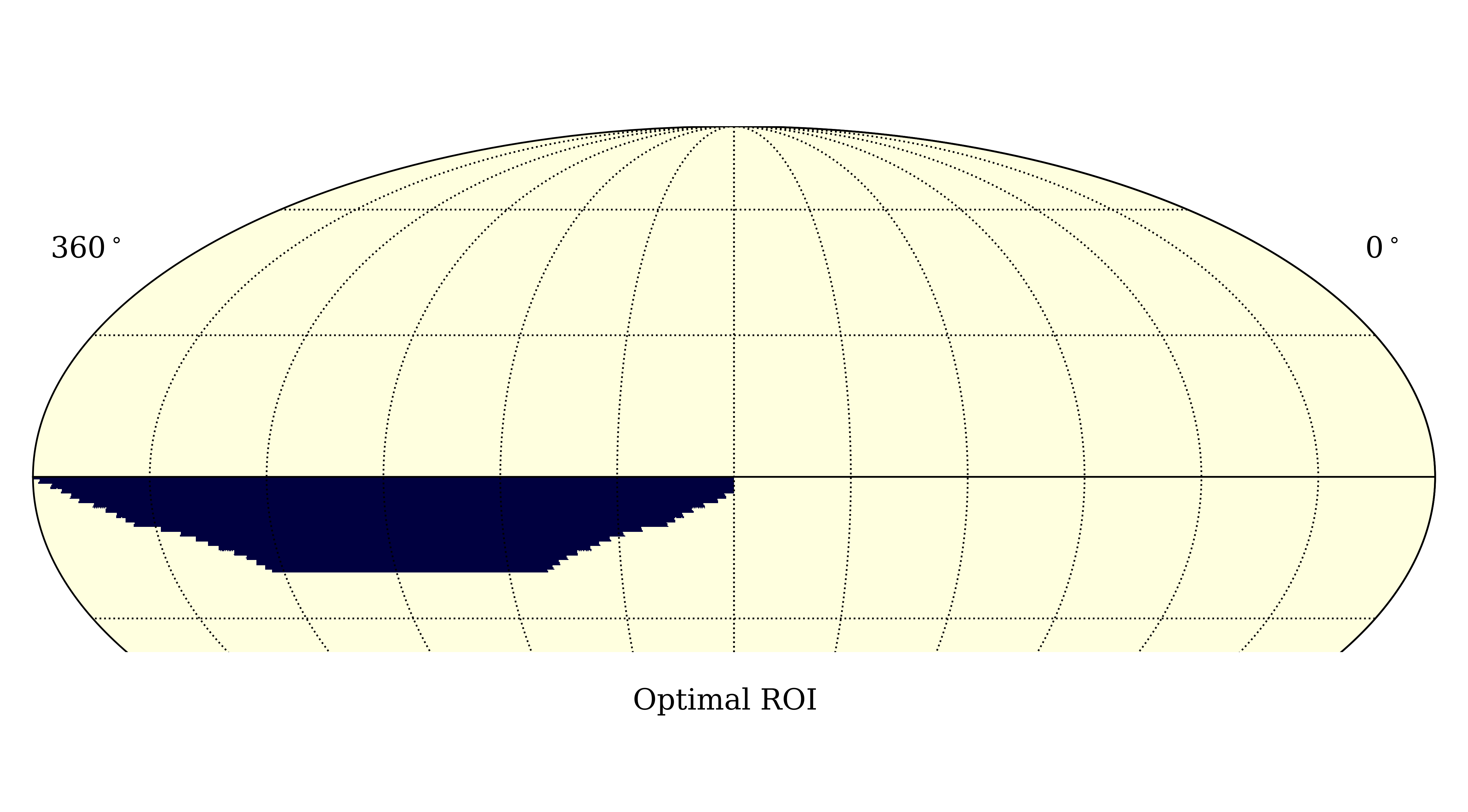}
	\caption{Optimal ROI chosen for this analysis based on the sensitivity plots in figure~\ref{fig:sensitivity_map}.  The ROI extends over declinations between 0 degrees and $-20$ degrees. To avoid contamination from normal-matter gamma-ray sources, known gamma-ray sources, as well as the extended emission on the north side of the Galactic Plane detected in ref.~\cite{pooja_thesis}, are removed during the search for dark matter.}
	\label{fig:hal-rois}
\end{figure}

Given the high number of detected HAWC sources associated with astrophysical emission in the Galactic plane, it is best to remove this region from the final ROI.  Additionally, all TeV-emitting sources with known astrophysical associations in the TeVCat catalog of known TeV sources \cite{tevcat} are removed to minimize possible contamination from sub-threshold sources. An additional constraint is made so the ROI excludes all pixels above zero degrees declination ($\mathrm{Dec}=0$) in order to avoid the new excess found in the $\alpha$-factor analysis \cite{pooja_thesis}.  Finally, we require a sensitivity cut such that the ROI at its widest spans no more than 180 degrees.  This final constraint is added to prevent the ROI from becoming arbitrarily wide and creating declinations where no pixels remain outside the ROI for use in the background calculation.  Under these constraints, both density profiles yield approximately the same optimal ROI, due to their shapes converging at distances far from the Galactic Halo center.
The chosen ROI that will be probed for Galactic dark matter emission, and also excluded from the background calculation, is displayed in figure~\ref{fig:hal-rois}.

\subsection{Source Modeling}
\label{subsec:model}

To compute the expected flux from such a large extended source, it is necessary to convolve the HAWC point spread function (PSF) and detector response with the source spatial profile.  
This is done using The HAWC Accelerated Likelihood (HAL) plugin \cite{hawchal,HAWC:2021lpf} within the Multi-Mission Maximum Likelihood (3ML) framework \cite{3ml,3mlgit}. To account for the varying HAWC detector response function across different declinations, the Galactic Halo ROI is broken into a set of smaller ROIs.  The convolution is then performed separately over each of these sub-ROIs and the resulting models linked to recover the full Galactic Halo extent.  To perform a fit, the likelihood approach summarized at the end of section~\ref{sec:hawc} is generalized by simply summing the log-likelihood contribution from each sub-ROI to obtain the full likelihood profile. 

In the event of a source as highly-extended as the Galactic Halo, the estimated background of the data maps will contain a substantial component of gamma rays from the dark matter itself. It is therefore only possible to resolve gamma-ray signals in excess of the background gamma-ray component, which will be referred to as effective flux (in contrast to the true total gamma-ray flux originating from hypothetical Galactic dark matter).  We estimate this effect by adding simulated dark matter emission from the Galactic Halo into the data and recomputing the $\alpha$-factor background (eq.~\ref{alpha}).  We then correct the expected excess in our fits by this factor, analogous to the effective flux corrections considered in the previous HAWC Galactic Halo analysis \cite{hawc_fb}.


To account for possible systematic uncertainties in the modeling of the HAWC detector, we consider the standard suite of detector uncertainties discussed in ref.~\cite{crab_high_energy}. The main source of systematic uncertainties within HAWC analyses come from discrepancies between the data and the simulated Monte Carlo events, which originate from uncertainties in the modeling of the detector, as detailed in ref.~\cite{crab_high_energy}.  

We consider both possible density profile models discussed in section~\ref{sec:gh}. As neither the Burkert (eq.~\ref{burkertgc}) or Einasto (eq.~\ref{einastogc}) profiles are strongly favored by current observations of the Galactic Center, both will be reported separately with neither being considered the nominal case.

\section{Results}
\label{sec:result}
\subsection{Fit results}
\label{subsec:fit}
With an ROI selected and known contamination removed, the results of the dark matter search are now shown.  We consider a set of representative annihilation and decay channels and a sample of dark matter masses between 10 and 100 TeV.  This mass range was chosen based on the dynamic range of the binning scheme used in this analysis as detailed in ref.~\cite{dwarf}.

\begin{figure}
	\includegraphics[width=0.5\textwidth]{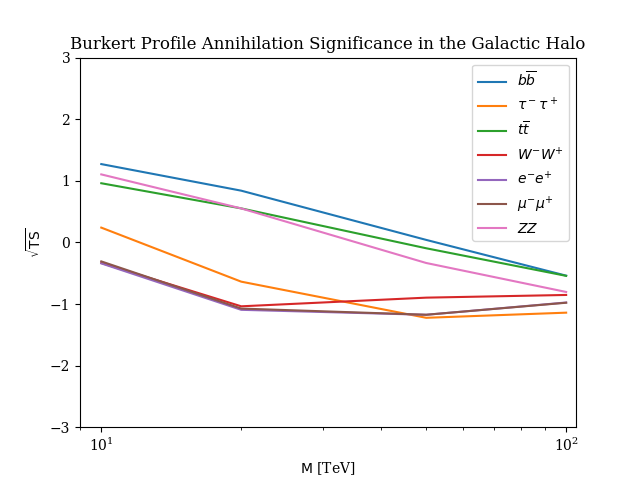}
	\includegraphics[width=0.5\textwidth]{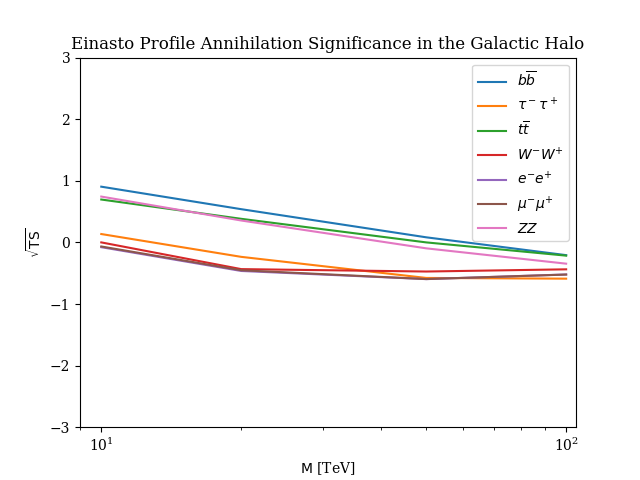}
	\includegraphics[width=0.5\textwidth]{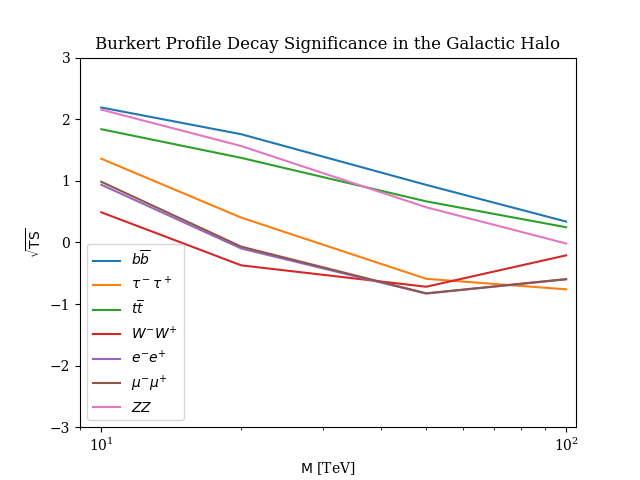}
	\includegraphics[width=0.5\textwidth]{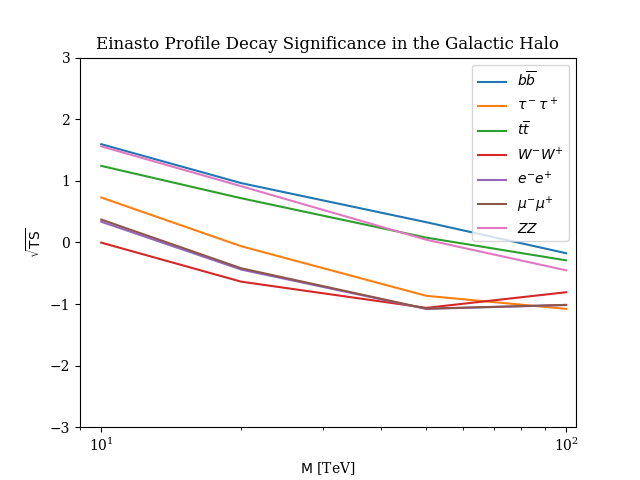}
	\caption{Significance ($\sqrt{\mathrm{TS}})$ as a function of dark matter mass and channel for annihilation assuming a Burkert profile (top left), annihilation assuming an Einasto profile (top right), decay assuming a Burkert profile (bottom left), and decay assuming an Einasto profile (bottom right).  No spectrum shows significant evidence of gamma-ray emission.}
	\label{fig:gh_dm_significance}
\end{figure}

The significance ($\sqrt{\mathrm{TS}})$ values for various spectra are plotted in figure~\ref{fig:gh_dm_significance}.  No significant emission is found for any spectrum.  Therefore, we set $95\%$ CL upper limits on the cross-section $\langle \sigma v \rangle$ and $95\%$ CL lower limits on the decay lifetime $\tau$.

\begin{figure}
	\includegraphics[width=0.5\textwidth]{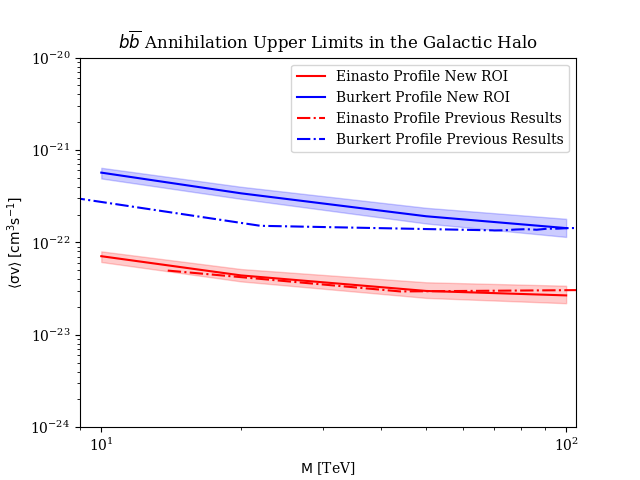}
	\includegraphics[width=0.5\textwidth]{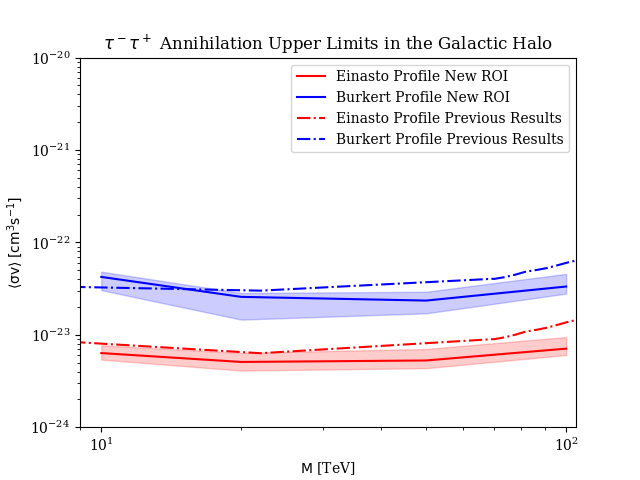}
	\includegraphics[width=0.5\textwidth]{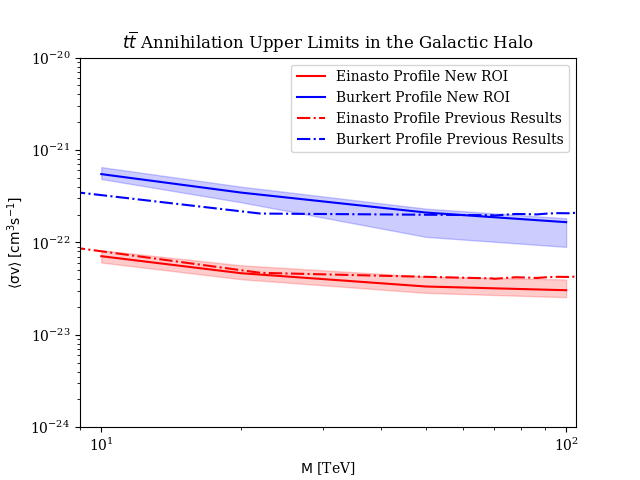}
	\includegraphics[width=0.5\textwidth]{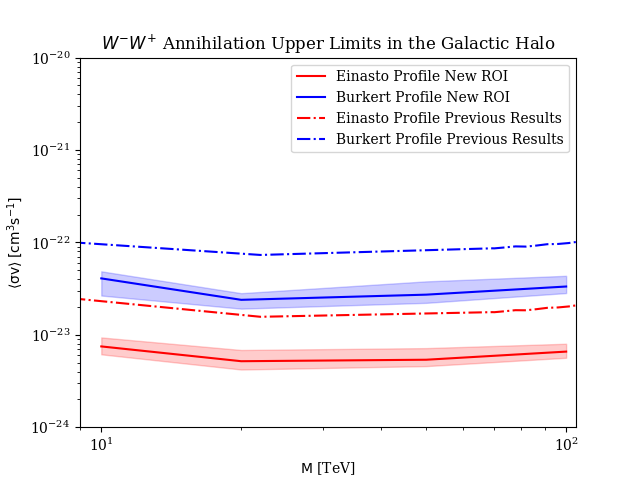}
	\caption{95\% CL upper limits on the velocity-weighted annihilation cross-section, $\langle \sigma v \rangle$, for the $b\overline{b}$ (top left), $\tau^+\tau^-$ (top right), $t\overline{t}$ (bottom left), and $W^+W^-$ (bottom right) dark matter spectra assuming the Einasto (red) and Burkert (blue) spatial profiles.  The correspondingly colored shaded regions are the systematic uncertainty bands.  The corresponding limits from the previous HAWC Galactic Halo analysis \cite{hawc_fb} are also plotted for comparison.}
	\label{fig:gh_upper_limits1}
\end{figure}

\begin{figure}
	\includegraphics[width=0.5\textwidth]{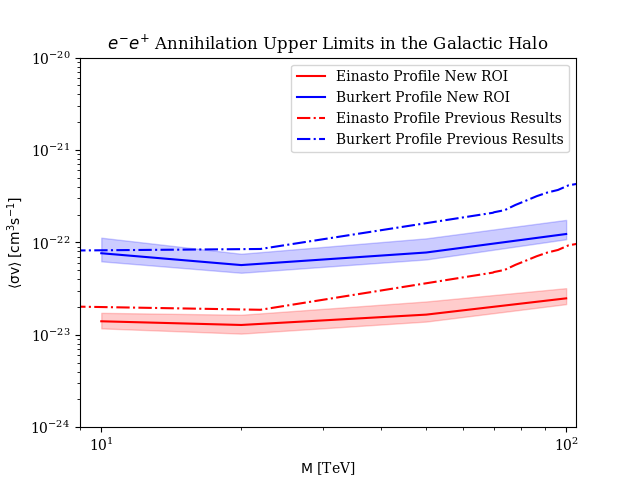}
	\includegraphics[width=0.5\textwidth]{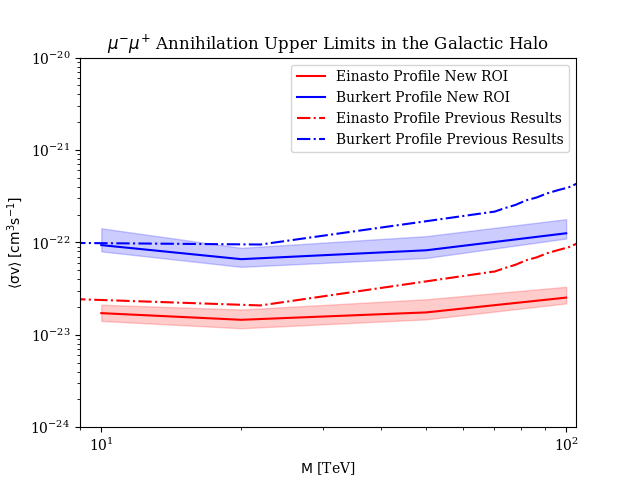}	\includegraphics[width=0.5\textwidth]{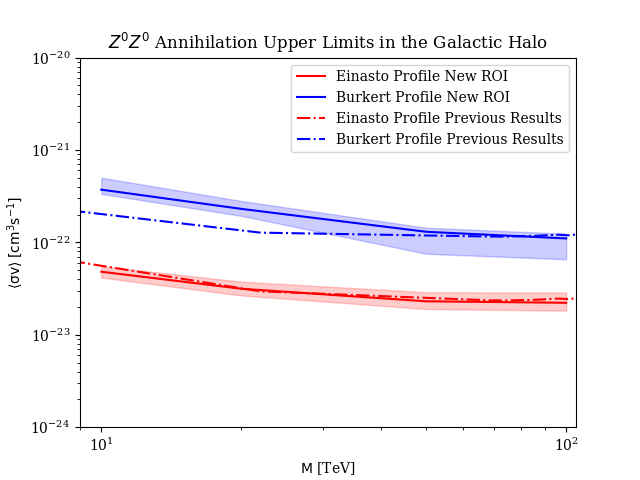}
	\caption{Additional 95\% CL upper limits on the velocity-weighted annihilation cross-section, $\langle \sigma v \rangle$, for the $e^+e^-$ (top left), $\mu^+\mu^-$ (top right), and $Z^0Z^0$ (bottom)  dark matter spectra assuming the Einasto (red) and Burkert (blue) spatial profiles.  The correspondingly colored shaded regions are the systematic uncertainty bands.  The corresponding limits from the previous HAWC Galactic Halo analysis \cite{hawc_fb} are also plotted for comparison.}
	\label{fig:gh_upper_limits2}
\end{figure}

\begin{figure}
	\includegraphics[width=0.5\textwidth]{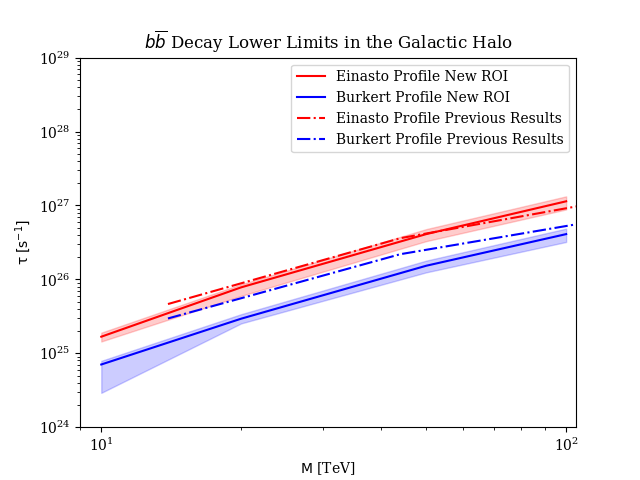}
	\includegraphics[width=0.5\textwidth]{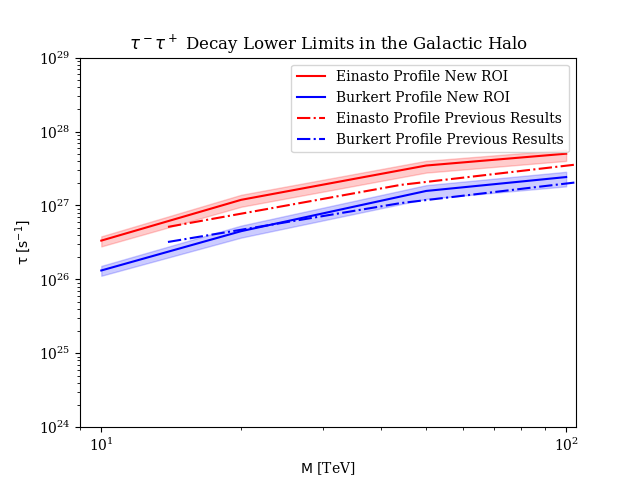}
	\includegraphics[width=0.5\textwidth]{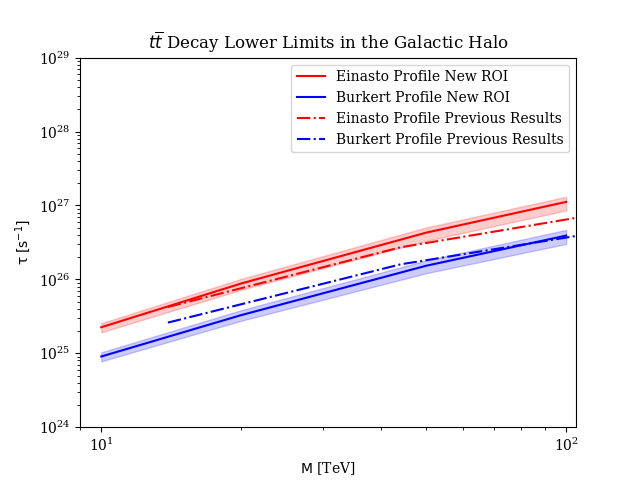}
	\includegraphics[width=0.5\textwidth]{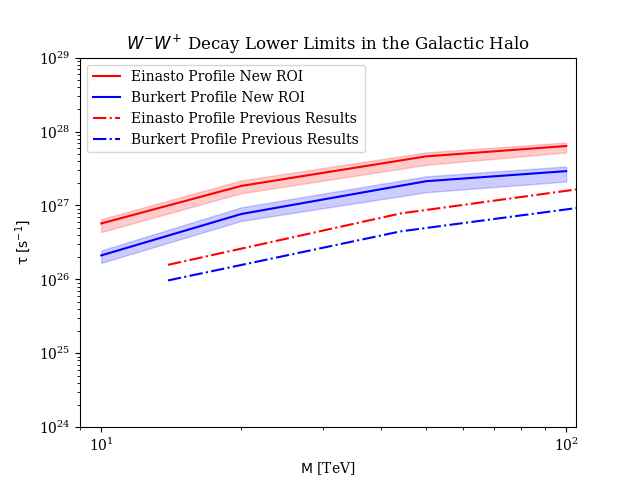}
	\caption{95\% CL lower limits on the decay lifetime, $\tau$, for the $b\overline{b}$ (top left), $\tau^+\tau^-$ (top right), $t\overline{t}$ (bottom left), and $W^+W^-$ (bottom right) dark matter spectra assuming the Einasto (red) and Burkert (blue) spatial profiles.  The correspondingly colored shaded regions are the systematic uncertainty bands.  The corresponding limits from the previous HAWC Galactic Halo analysis \cite{hawc_fb} are also plotted for comparison.}
	\label{fig:gh_lower_limits1}
\end{figure}

\begin{figure}
	\includegraphics[width=0.5\textwidth]{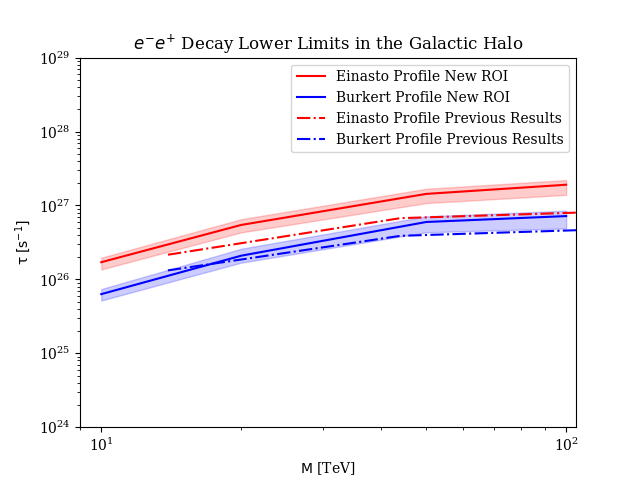}
	\includegraphics[width=0.5\textwidth]{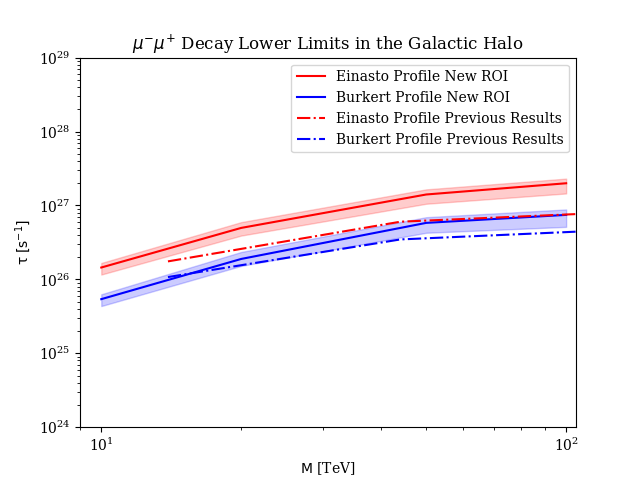}
	\includegraphics[width=0.5\textwidth]{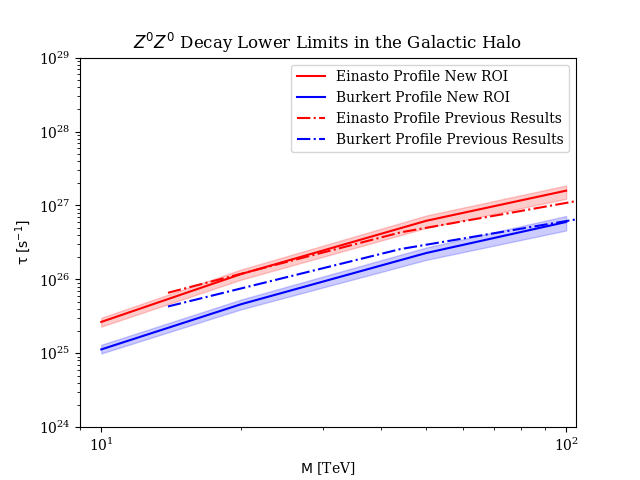}
	\caption{Additional 95\% CL lower limits on the decay lifetime, $\tau$, for the $e^+e^-$ (top left), $\mu^+\mu^-$ (top right), and $Z^0Z^0$ (bottom) dark matter spectra assuming the Einasto (red) and Burkert (blue) spatial profiles. The correspondingly colored shaded regions are the systematic uncertainty bands. The corresponding limits from the previous HAWC Galactic Halo analysis \cite{hawc_fb} are also plotted for comparison.}
	\label{fig:gh_lower_limits2}
\end{figure}

As described in section~\ref{sec:hawc}, the $\mathrm{TS}$ (eq.~\ref{ts}) quantity asymptotically follows a $\chi$-squared distribution with one degree of freedom.  The limits are then set by increasing the free parameter corresponding to flux normalization (the cross-section, $\langle \sigma v \rangle$, for annihilation and reciprocal lifetime, $\frac{1}{\tau}$, for decay) from the maximum-likelihood value until the likelihood ratio decreases by an appropriate amount for a given confidence level.  To obtain 95\% CL limits, 1.355  is chosen as the appropriate value (see the Appendix of ref.~\cite{dwarf} for details). For annihilation this results in upper limits on $\langle \sigma v \rangle$, while for decay this corresponds to a lower limits on $\tau$.  The annihilation upper limits are shown in figure~\ref{fig:gh_upper_limits1} and figure~\ref{fig:gh_upper_limits2}, while the decay lower limits are shown in figure~\ref{fig:gh_lower_limits1} and figure~\ref{fig:gh_lower_limits2}.  

As expected, the density profile choice has a substantial effect on the limits.  However, the difference is only on the order of a factor of 2--3, rather than the many orders of magnitude expected if the Galactic Center were to be the only part of the halo considered.  Neither choice of profile is treated as the nominal case here, so all systematics are computed independently for the limits arising in both cases. The uncertainty bands in figures~\ref{fig:gh_upper_limits1}, \ref{fig:gh_upper_limits2}, \ref{fig:gh_lower_limits1}, and \ref{fig:gh_lower_limits2} are computed by adding the contribution from each systematic in quadrature. This quadrature addition is done separately for the positive and negative halves of the bands, resulting in an effect of about 11\% on the positive side and 23\% on the negative side.

The limits from the previous HAWC analysis (ref.\cite{hawc_fb}) are also plotted in figures~\ref{fig:gh_upper_limits1}, \ref{fig:gh_upper_limits2}, \ref{fig:gh_lower_limits1}, and \ref{fig:gh_lower_limits2}. The current results show a strong improvement for the $W^+W^-$ channel due to the inclusion of electroweak corrections, while the other channels show consistent improvement in the higher masses.  At the lower masses, the previous results appear to produce stronger limits in certain channels due to statistical fluctuations.  As can be seen in Fig of ref.~\cite{hawc_fb}, the results of the previous analysis were heavily influenced by strong downward statistical fluctuations that are not present in the current analysis.  This effect is discussed further in section~\ref{subsec:stats}.



\subsection{Statistical effects}
\label{subsec:stats}

To compute the effects of statistical fluctuations on the limits, 500 pseudomaps are generated, where the data in each bin is replaced with a randomly drawn value from a Poisson distribution with a mean equal to the estimated background in that bin. The 95\% upper and lower limits are recomputed using these pseudomaps in place of the actual data.  The median of the resulting distribution results in ``expected" limits; i.e., the limits one would expect assuming the null hypothesis of a background-only model is true.  The 68\% and 95\% containment bands are computed about the null hypothesis.  Assuming the background was well-modeled in this analysis, the limits should be no further from the expected case than the 95\% band, being influenced only by Poisson fluctuations and not an undetected source or mismodeled background.  

\begin{figure}
	\includegraphics[width=0.5\textwidth]{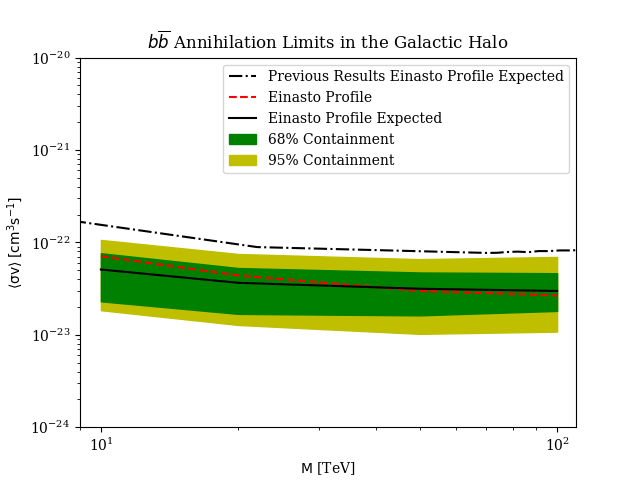}
	\includegraphics[width=0.5\textwidth]{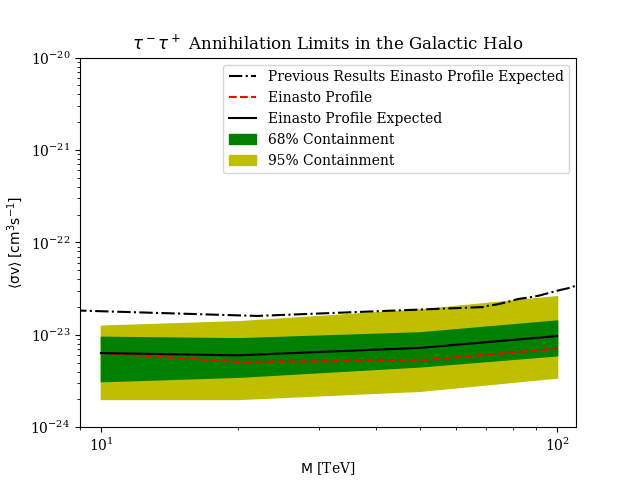}
	\caption{Statistical bands for 95\% CL upper limits on the dark matter annihilation cross- section,$\langle \sigma v \rangle$, for the $b\overline{b}$ (left) and $\tau^+\tau^-$ (right) channels assuming an Einasto spatial profile.  In both cases, the actual limits fall well within the expected 95\% range from statistical fluctuations of the null hypothesis.  The expected limits from the previous HAWC analysis \cite{hawc_fb} are also plotted to more clearly show the improvement in sensitivity from this analysis independent of statistical fluctuation effects.}
	\label{fig:einasto_brazilband_annihilation}
\end{figure}

\begin{figure}
	\includegraphics[width=0.5\textwidth]{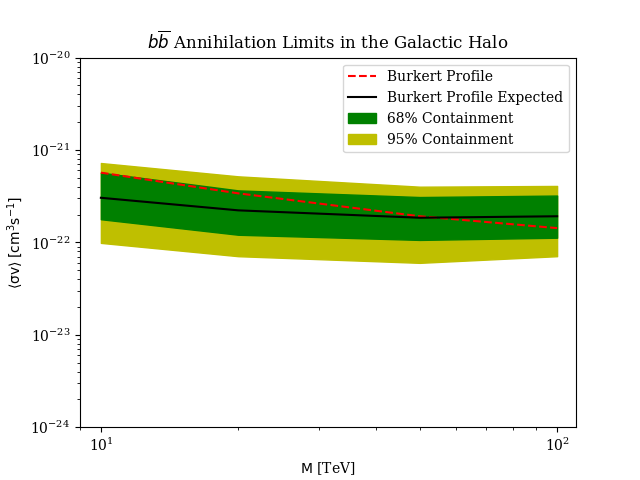}
	\includegraphics[width=0.5\textwidth]{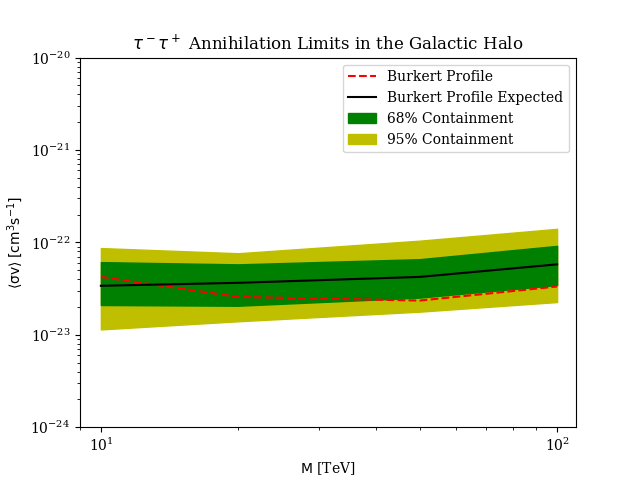}
	\caption{Statistical bands for 95\% CL upper limits on the dark matter annihilation cross-section,$\langle \sigma v \rangle$, for the $b\overline{b}$ (left) and $\tau^+\tau^-$ (right) channels assuming a Burkert spatial profile.  In both cases, the actual limits fall well within the expected 95\% range from statistical fluctuations of the null hypothesis.}
	\label{fig:burkert_brazilband_annihilation}
\end{figure}

\begin{figure}
	\includegraphics[width=0.5\textwidth]{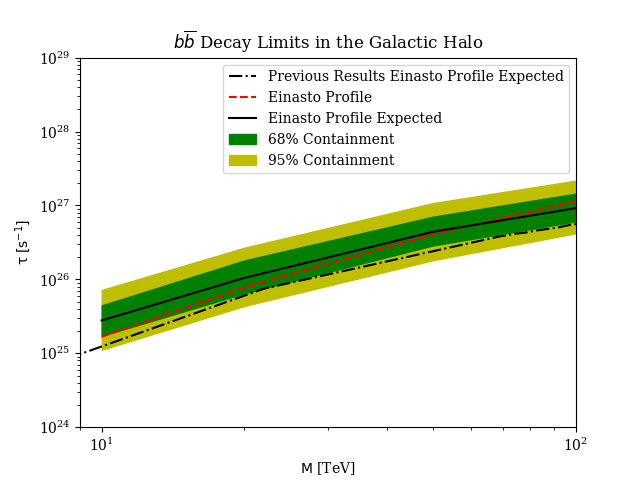}
	\includegraphics[width=0.5\textwidth]{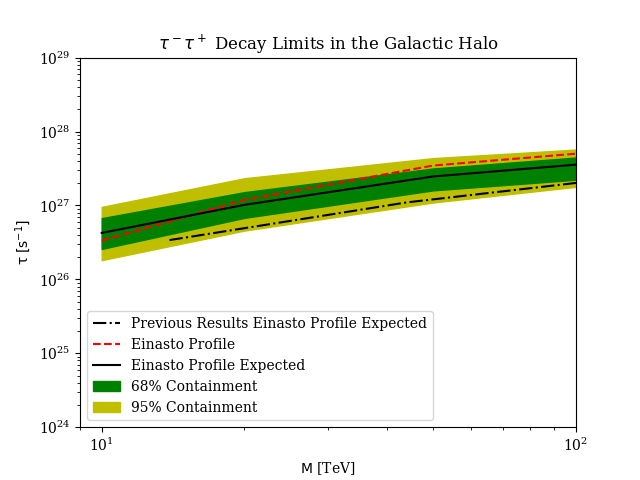}
	\caption{Statistical bands for 95\% CL lower limits on the dark matter decay lifetime, $\tau$, for the $b\overline{b}$ (left) and $\tau^+\tau^-$ (right) channels assuming an Einasto spatial profile.  In both cases, the actual limits fall well within the expected 95\% range from statistical fluctuations of the null hypothesis.  The expected limits from the previous HAWC analysis \cite{hawc_fb} are also plotted to more clearly show the improvement in sensitivity from this analysis independent of statistical fluctuation effects.}
	\label{fig:einasto_brazilband_decay}
\end{figure}

\begin{figure}
	\includegraphics[width=0.5\textwidth]{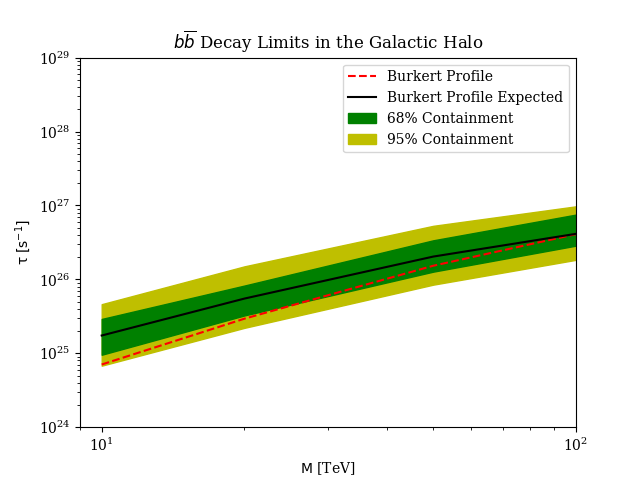}
	\includegraphics[width=0.5\textwidth]{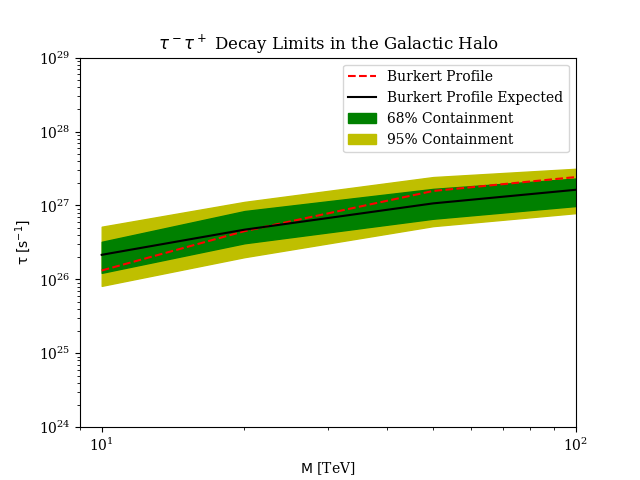}
	\caption{Statistical bands for 95\% CL lower limits on the dark matter decay lifetime, $\tau$, for the $b\overline{b}$ (left) and $\tau^+\tau^-$ (right) channels assuming a Burkert spatial profile.  In both cases, the actual limits fall well within the expected 95\% range from statistical fluctuations of the null hypothesis.}
	\label{fig:burkert_brazilband_decay}
\end{figure}

As this process is computationally intense, the statistical bands for only two sample spectra, $b\overline{b}$ and $\tau^+\tau^-$, are plotted for all masses and spatial profiles.  These are, respectively, the softest (most low-energy-dominated) and hardest (most high-energy-dominated) channels and therefore span the extremes of the channel parameter space.  The results are plotted for annihilation and decay, as well as the two spatial profiles considered, in figures~\ref{fig:einasto_brazilband_annihilation},~\ref{fig:burkert_brazilband_annihilation},~\ref{fig:einasto_brazilband_decay}, and \ref{fig:burkert_brazilband_decay}.  The expected limits from the previous HAWC analysis are also plotted for comparison when available (figures~\ref{fig:einasto_brazilband_annihilation} and \ref{fig:einasto_brazilband_decay}).  Comparing the expected limits more clearly shows the improvement in sensitivity gained from the current analysis by removing the effect of statistical fluctuations.  All of the plotted limits for this analysis fall well within the 95\% band and are highly consistent with expectation.  Therefore, the results presented here are more robust to statistical effects than those published in ref.~\cite{hawc_fb}.

\begin{figure}
	\includegraphics[width=0.5\textwidth]{figures/galactic-halo-bb-einasto-annihilation-nonprofile-eq-sourcemask3-limits-brazil-band.png}
	\includegraphics[width=0.5\textwidth]{figures/galactic-halo-tautau-einasto-annihilation-nonprofile-eq-sourcemask3-limits-brazil-band.png}
	\includegraphics[width=0.5\textwidth]{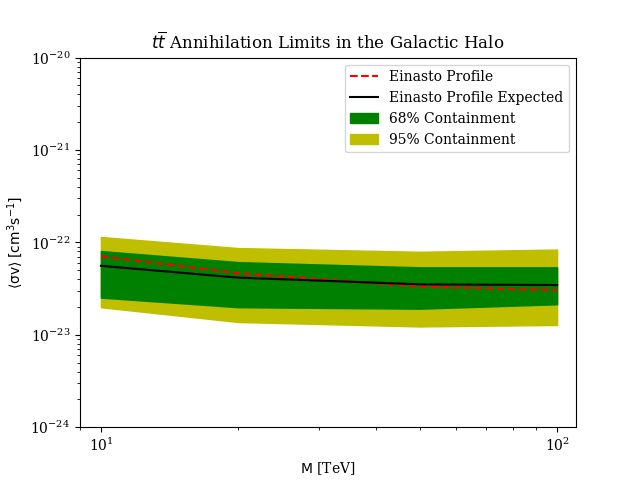}
	\includegraphics[width=0.5\textwidth]{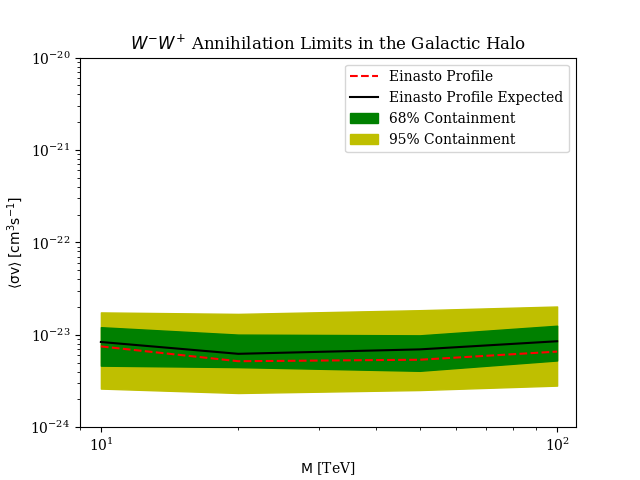}
	\caption{Statistical bands on the 95\% CL upper limits on the velocity-weighted annihilation cross-section, $\langle \sigma v \rangle$, for the $b\overline{b}$ (top left), $\tau^+\tau^-$ (top right), $t\overline{t}$ (bottom left), and $W^+W^-$ (bottom right) dark matter spectra assuming an Einasto profile.  This is a supplementary figure meant to demonstrate the all channels are within the 95\% CL statistical uncertainty range for a null test case.}
	\label{fig:complete_einasto_brazil_band1}
\end{figure}

\begin{figure}
	\includegraphics[width=0.5\textwidth]{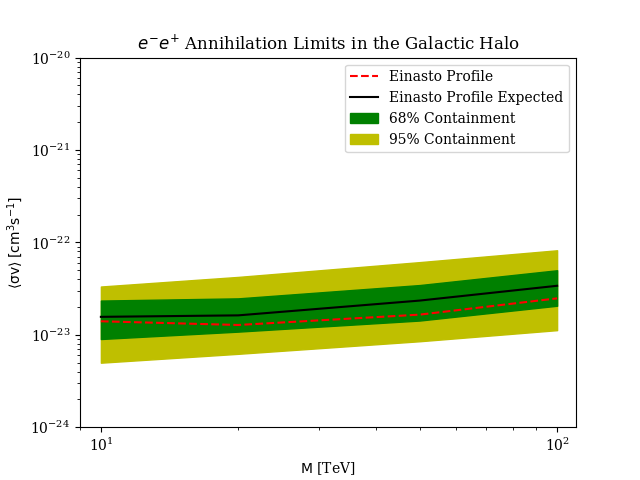}
	\includegraphics[width=0.5\textwidth]{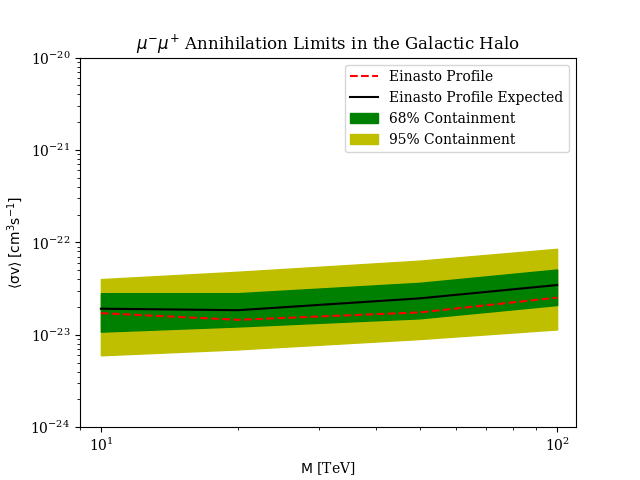}
	\includegraphics[width=0.5\textwidth]{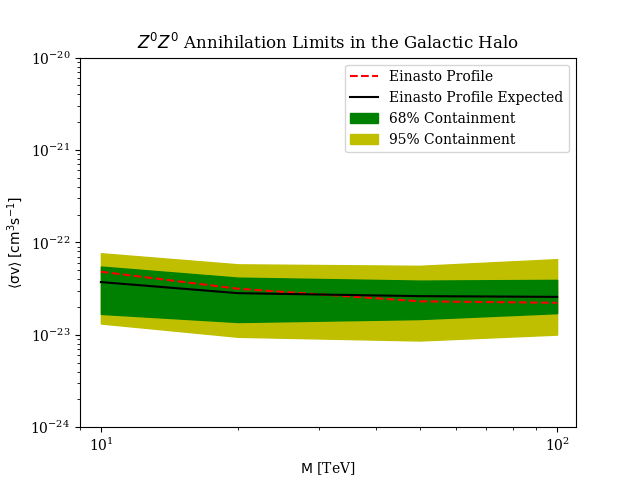}
	\caption{Additional statistical bands on the 95\% CL upper limits on the velocity-weighted annihilation cross-section ,$\langle \sigma v \rangle$, for the $e^+e^-$ (top left), $\mu^+\mu^-$ (top right), and $Z^0Z^0$ (bottom) dark matter spectra assuming an Einasto profile.  This is a supplementary figure meant to demonstrate the all channels are within the 95\% CL statistical uncertainty range for a null test case.}
	\label{fig:complete_einasto_brazil_band2}
\end{figure}

To further verify the limits are within expectation true across all channels, a complete set of plots with statistical bands for all channels was made for the Einasto profile annihilation case and is shown in figures~\ref{fig:complete_einasto_brazil_band1} and \ref{fig:complete_einasto_brazil_band2}.  The limits in these plots are also well within the 95\% bands.

\section{Conclusion}
\label{sec:conclustion}

 Although no significant evidence of dark matter gamma-ray emission was found, the new constraints on dark matter annihilation and decay derived from this analysis show a marked improvement over the previous published results in ref.~\cite{hawc_fb}.  Additionally, the inclusion of the electroweak corrections dramatically improves the results for the $W^+ W^-$ channel due to the higher relative flux at the highest energies.  Furthermore, the background is now better estimated and the results are not biased toward negative flux as was seen in ref.~\cite{hawc_fb}. As expected, the limits are relatively robust against the
 large uncertainties arising from the density profile choice.  Rather than the multi-order-of-magnitude difference that one would expect if only the Galactic Center were considered (see figure~\ref{fig:profiles-comparison}), these results differ by less than a single order of magnitude between the profiles.  
 
 These results are therefore able to constrain dark matter gamma-ray emission from the Galactic Halo for a wide range of possible density profile behaviors and do not rely on the assumption of a strong central cusp.  Current evidence from both observation and N-body simulations that include baryonic components favors profiles that lack such a cusp, necessitating the approach used to observe emission from a larger region of the Halo utilized in this analysis \cite{cusp_flatteing}.  Therefore, these results show the possibility of constraining dark matter even in the less optimistic case favored by current astrophysical data.  Future work from both HAWC and other wide-field-of-view experiments will build on these results and continue to probe dark matter from the Galactic Halo, even if the exact density profile behavior remains unknown.


\clearpage
\acknowledgments

We acknowledge the support from: the US National Science Foundation (NSF); the US Department of Energy Office of High-Energy Physics; 
the Laboratory Directed Research and Development (LDRD) program of Los Alamos National Laboratory; 
Consejo Nacional de Ciencia y Tecnolog\'{\i}a (CONACyT), M{\'e}xico (grants 271051, 232656, 260378, 179588, 239762, 254964, 271737, 258865, 243290, 132197, 281653)(C{\'a}tedras 873, 1563, 341), Laboratorio Nacional HAWC de rayos gamma; 
L'OREAL Fellowship for Women in Science 2014; 
Red HAWC, M{\'e}xico; 
DGAPA-UNAM (grants IG100317, IN111315, IN111716-3, IA102715, IN109916, IA102917, IN112218); 
VIEP-BUAP; 
PIFI 2012, 2013, PROFOCIE 2014, 2015; 
the University of Wisconsin Alumni Research Foundation; 
the Institute of Geophysics, Planetary Physics, and Signatures at Los Alamos National Laboratory; 
Polish Science Centre grant DEC-2014/13/B/ST9/945, DEC-2017/27/B/ST9/02272; 
Coordinaci{\'o}n de la Investigaci{\'o}n Cient\'{\i}fica de la Universidad Michoacana; Royal Society - Newton Advanced Fellowship 180385. Thanks to Scott Delay, Luciano D\'{\i}az and Eduardo Murrieta for technical support.
\bibliography{gh_dm_bibliography}

\providecommand{\href}[2]{#2}\begingroup\raggedright\begin{thebibliography}{10}

\bibitem{andernach2017english}
H.~Andernach and F.~Zwicky, {\it English and spanish translation of zwicky's
  (1933) the redshift of extragalactic nebulae},  2017.

\bibitem{dark_evidence}
G.~Bertone, D.~Hooper, and J.~Silk, {\it {Particle dark matter: Evidence,
  candidates and constraints}},  {\em Phys. Rept.} {\bf 405} (2005) 279--390,
  [\href{http://arxiv.org/abs/hep-ph/0404175}{{\tt hep-ph/0404175}}].

\bibitem{wimp}
G.~Jungman, M.~Kamionkowski, and K.~Griest, {\it {Supersymmetric dark matter}},
   {\em Phys. Rept.} {\bf 267} (1996) 195--373,
  [\href{http://arxiv.org/abs/hep-ph/9506380}{{\tt hep-ph/9506380}}].

\bibitem{planck}
A.~A. Dutton and A.~V. Macci{\`o}, {\it Cold dark matter haloes in the planck
  era: evolution of structural parameters for einasto and nfw profiles},  {\em
  Monthly Notices of the Royal Astronomical Society} {\bf 441} (2014), no.~4
  3359--3374.

\bibitem{wimpproceeding}
M.~Kamionkowski, {\it {WIMP and axion dark matter}},  in {\em {ICTP Summer
  School in High-Energy Physics and Cosmology}}, pp.~394--411, 6, 1997.
\newblock \href{http://arxiv.org/abs/hep-ph/9710467}{{\tt hep-ph/9710467}}.

\bibitem{dark_back}
K.~N. Abazajian and J.~P. Harding, {\it {Constraints on WIMP and
  Sommerfeld-Enhanced Dark Matter Annihilation from Observations of the
  Galactic Center}},  {\em JCAP} {\bf 1201} (2012) 041,
  [\href{http://arxiv.org/abs/1110.6151}{{\tt arXiv:1110.6151}}].

\bibitem{patproceeding}
J.~P. Harding, {\it Dark matter indirect detection with gamma rays},  {\em
  Journal of Physics: Conference Series} {\bf 866} (jun, 2017) 012007.

\bibitem{pppc}
M.~Cirelli, G.~Corcella, A.~Hektor, G.~Hutsi, M.~Kadastik, P.~Panci, M.~Raidal,
  F.~Sala, and A.~Strumia, {\it {PPPC 4 DM ID: A Poor Particle Physicist
  Cookbook for Dark Matter Indirect Detection}},  {\em JCAP} {\bf 03} (2011)
  051, [\href{http://arxiv.org/abs/1012.4515}{{\tt arXiv:1012.4515}}].
  [Erratum: JCAP 10, E01 (2012)].

\bibitem{einasto3}
J.~{Einasto}, {\it {On the Construction of a Composite Model for the Galaxy and
  on the Determination of the System of Galactic Parameters}},  {\em Trudy
  Astrofizicheskogo Instituta Alma-Ata} {\bf 5} (Jan., 1965) 87--100.

\bibitem{einasto1}
J.~Stadel, D.~Potter, B.~Moore, J.~Diemand, P.~Madau, M.~Zemp, M.~Kuhlen, and
  V.~Quilis, {\it {Quantifying the heart of darkness with GHALO - a
  multi-billion particle simulation of our galactic halo}},  {\em Mon. Not.
  Roy. Astron. Soc.} {\bf 398} (2009) L21--L25,
  [\href{http://arxiv.org/abs/0808.2981}{{\tt arXiv:0808.2981}}].

\bibitem{einasto2}
J.~F. Navarro, A.~Ludlow, V.~Springel, J.~Wang, M.~Vogelsberger, S.~D.~M.
  White, A.~Jenkins, C.~S. Frenk, and A.~Helmi, {\it {The Diversity and
  Similarity of Cold Dark Matter Halos}},  {\em Mon. Not. Roy. Astron. Soc.}
  {\bf 402} (2010) 21, [\href{http://arxiv.org/abs/0810.1522}{{\tt
  arXiv:0810.1522}}].

\bibitem{clumpy}
V.~Bonnivard, M.~Hütten, E.~Nezri, A.~Charbonnier, C.~Combet, and D.~Maurin,
  {\it {CLUMPY : Jeans analysis, gamma-ray and neutrino fluxes from dark matter
  (sub-)structures}},  {\em Comput. Phys. Commun.} {\bf 200} (2016) 336--349,
  [\href{http://arxiv.org/abs/1506.07628}{{\tt arXiv:1506.07628}}].

\bibitem{burkert}
A.~Burkert, {\it {The Structure of dark matter halos in dwarf galaxies}},  {\em
  IAU Symp.} {\bf 171} (1996) 175,
  [\href{http://arxiv.org/abs/astro-ph/9504041}{{\tt astro-ph/9504041}}].
  [Astrophys. J.447,L25(1995)].

\bibitem{cusp_flatteing}
C.~Nipoti and J.~Binney, {\it Early flattening of dark matter cusps in dwarf
  spheroidal galaxies},  {\em Monthly Notices of the Royal Astronomical
  Society} {\bf 446} (Nov, 2014) 1820–1828.

\bibitem{Navarro:2008kc}
J.~F. Navarro, A.~Ludlow, V.~Springel, J.~Wang, M.~Vogelsberger, S.~D.~M.
  White, A.~Jenkins, C.~S. Frenk, and A.~Helmi, {\it {The Diversity and
  Similarity of Cold Dark Matter Halos}},  {\em Mon. Not. Roy. Astron. Soc.}
  {\bf 402} (2010) 21, [\href{http://arxiv.org/abs/0810.1522}{{\tt
  arXiv:0810.1522}}].

\bibitem{HAWC:2020hrt}
{\bf HAWC} Collaboration, A.~Albert et~al., {\it {3HWC: The Third HAWC Catalog
  of Very-High-Energy Gamma-ray Sources}},  {\em Astrophys. J.} {\bf 905}
  (2020), no.~1 76, [\href{http://arxiv.org/abs/2007.08582}{{\tt
  arXiv:2007.08582}}].

\bibitem{hawcback}
A.~U. Abeysekara et~al., {\it {Observation of the Crab Nebula with the HAWC
  Gamma-Ray Observatory}},  {\em Astrophys. J.} {\bf 843} (2017), no.~1 39,
  [\href{http://arxiv.org/abs/1701.01778}{{\tt arXiv:1701.01778}}].

\bibitem{historical:2023opo}
{\bf historical, present HAWC} Collaboration, A.~U. Abeysekara et~al., {\it
  {The High-Altitude Water Cherenkov (HAWC) observatory in M\'exico: The
  primary detector}},  {\em Nucl. Instrum. Meth. A} {\bf 1052} (2023) 168253,
  [\href{http://arxiv.org/abs/2304.00730}{{\tt arXiv:2304.00730}}].

\bibitem{healpix}
K.~M. Gorski, E.~Hivon, A.~J. Banday, B.~D. Wandelt, F.~K. Hansen, M.~Reinecke,
  and M.~Bartelman, {\it {HEALPix - A Framework for high resolution
  discretization, and fast analysis of data distributed on the sphere}},  {\em
  Astrophys. J.} {\bf 622} (2005) 759--771,
  [\href{http://arxiv.org/abs/astro-ph/0409513}{{\tt astro-ph/0409513}}].

\bibitem{dwarf}
A.~Albert et~al., {\it {Dark Matter Limits From Dwarf Spheroidal Galaxies with
  The HAWC Gamma-Ray Observatory}},  {\em Astrophys. J.} {\bf 853} (2018),
  no.~2 154, [\href{http://arxiv.org/abs/1706.01277}{{\tt arXiv:1706.01277}}].

\bibitem{wilks}
S.~S. Wilks, {\it The large-sample distribution of the likelihood ratio for
  testing composite hypotheses},  {\em Ann. Math. Statist.} {\bf 9} (03, 1938)
  60--62.

\bibitem{pooja_thesis}
P.~Surajbali, {\em Observing large-scale structures in the gamma-ray sky}.
\newblock PhD thesis, Ruprecht Karl University of Heidelberg, Heidelberg,
  Germany, 7, 2020.
\newblock https://archiv.ub.uni-heidelberg.de/volltextserver/28625/.

\bibitem{li-ma}
T.-P. Li and Y.-Q. Ma, {\it Analysis methods for results in gamma-ray
  astronomy},  {\em The Astrophysical Journal} {\bf 272} (1983) 317--324.

\bibitem{all-sky}
{\bf HAWC} Collaboration, A.~U. Abeysekara et~al., {\it {Searching for Dark
  Matter Sub-structure with HAWC}},  {\em JCAP} {\bf 1907} (2019), no.~07 022,
  [\href{http://arxiv.org/abs/1811.11732}{{\tt arXiv:1811.11732}}].

\bibitem{tevcat}
TeVCat homepage: http://tevcat.uchicago.edu/.

\bibitem{hawchal}
G.~Vianello, ``Hawc accelerated likelihood.''
  https://github.com/threeML/hawc{\_}hal/, 2018.

\bibitem{HAWC:2021lpf}
{\bf HAWC} Collaboration, A.~U. Abeysekara et~al., {\it {Characterizing
  gamma-ray sources with HAL (HAWC Accelerated likelihood) and 3ML}},  {\em
  PoS} {\bf ICRC2021} (2021) 828.

\bibitem{3ml}
G.~Vianello, R.~J. Lauer, P.~Younk, L.~Tibaldo, J.~M. Burgess, H.~Ayala,
  P.~Harding, M.~Hui, N.~Omodei, and H.~Zhou, {\it {The Multi-Mission Maximum
  Likelihood framework (3ML)}},  2015.
\newblock \href{http://arxiv.org/abs/1507.08343}{{\tt arXiv:1507.08343}}.

\bibitem{3mlgit}
G.~Vianello, ``3ml.'' https://github.com/threeML/threeML, 2016.

\bibitem{hawc_fb}
{\bf HAWC} Collaboration, A.~Abeysekara et~al., {\it {A Search for Dark Matter
  in the Galactic Halo with HAWC}},  {\em JCAP} {\bf 02} (2018) 049,
  [\href{http://arxiv.org/abs/1710.10288}{{\tt arXiv:1710.10288}}].

\bibitem{crab_high_energy}
{\bf HAWC} Collaboration, A.~Abeysekara et~al., {\it {Measurement of the Crab
  Nebula at the Highest Energies with HAWC}},  {\em Astrophys. J.} {\bf 881}
  (2019) 134, [\href{http://arxiv.org/abs/1905.12518}{{\tt arXiv:1905.12518}}].

\end{thebibliography}\endgroup

\end{document}